\documentstyle[12pt,aaspp4,epsfig]{article}

\def \ref {\noindent\hangindent=1.0in\hangafter=1}

\def\ltsima{$\; \buildrel < \over \sim \;$}
\def\simlt{\lower.5ex\hbox{\ltsima}} 
\def\gtsima{$\; \buildrel > \over \sim \;$}
\def\simgt{\lower.5ex\hbox{\gtsima}} 

\begin{document}

\title{Ultraviolet and Multiwavelength Variability\\
of the Blazar 3C~279: Evidence for Thermal Emission}

\author{E. Pian\altaffilmark{1,2,3}, 
C. M. Urry\altaffilmark{1,3}, 
L. Maraschi\altaffilmark{4}, 
G. Madejski\altaffilmark{5}, 
I. M. McHardy\altaffilmark{6},
A. Koratkar\altaffilmark{1},}
\author{A. Treves\altaffilmark{7},
L. Chiappetti\altaffilmark{8}, 
P. Grandi\altaffilmark{9,3},
R. C. Hartman\altaffilmark{5},
H. Kubo\altaffilmark{10},
C. M. Leach\altaffilmark{6},}
\author{J. E. Pesce\altaffilmark{1},
C. Imhoff\altaffilmark{1,11},
R. Thompson\altaffilmark{1,11}, 
A. E. Wehrle\altaffilmark{12}} 

\altaffiltext{1}{Space Telescope Science Institute, 3700 San Martin
Drive, Baltimore, MD 21218}
\altaffiltext{2}{Present address: Istituto di Tecnologie e Studio delle
Radiazioni
Extraterrestri, CNR, Via Gobetti 101, I-40129 Bologna, Italy}
\altaffiltext{3}{Guest Observer with the {\it International Ultraviolet
Explorer}}
\altaffiltext{4}{Osservatorio Astronomico di Brera, via Brera 28,
I-20121 Milan, Italy}
\altaffiltext{5}{Laboratory for High Energy Astrophysics, Goddard Space 
Flight Center, Greenbelt, MD 20771}
\altaffiltext{6}{Department of Physics, University of Southampton,
Southampton SO9 5NH, UK}
\altaffiltext{7}{Department of Physics, University of Como, Via
Lucini 3, I-22100 Como, Italy}
\altaffiltext{8}{Istituto di Fisica Cosmica e Tecnologie Relative, CNR, 
via Bassini 15, I-20133 Milan, Italy}
\altaffiltext{9}{Istituto di Astrofisica Spaziale, CNR, via
Fosso del 
Cavaliere, Area di Ricerca Tor Vergata, I-00133 Rome, Italy}
\altaffiltext{10}{Department of Physics, Tokyo Institute of 
Technology, 2-12-1 Ookayama, Meguro, Tokyo 152-8551, Japan}
\altaffiltext{11}{Science Programs, Computer Sciences Corp., 1100 West Street,
Laurel, MD 20707}
\altaffiltext{12}{Infrared Processing Analysis Center, MC
100-22, Jet Propulsion Laboratory and California
Institute of Technology, Pasadena, CA 91125}

\begin{abstract}

The $\gamma$-ray blazar 3C~279 was monitored on a nearly daily basis with
IUE, ROSAT and EGRET for three weeks between December 1992 and January
1993. During this period, the blazar was at a historical minimum at all
wavelengths.  Here we present the
UV data obtained during the above multiwavelength campaign.  A maximum UV
variation of $\sim$50\% is detected, while during the same period the
X-ray flux varied by no more than 13\%. 

At the lowest UV flux level the average spectrum in the 1230-2700 \AA\
interval is unusually flat for this object ($\langle \alpha_{UV} \rangle
\sim 1$). The flattening could represent the lowest energy tail of the
inverse Compton component responsible for the X-ray emission, or could be
due to the presence of a thermal component at $\sim$20000~K possibly
associated with an accretion disk.  The presence of an accretion disk in
this blazar object, likely observable only in very low states and
otherwise hidden by the beamed, variable synchrotron component, would be
consistent with the scenario in which the seed photons for the inverse
Compton mechanism producing the $\gamma$-rays are external to the
relativistic jet. 

We further discuss the long term correlation of the UV flux with the
X-ray and $\gamma$-ray fluxes obtained at various epochs.  All UV
archival data are included in the analysis.  Both the X- and $\gamma$-ray
fluxes are generally well correlated with the UV flux, approximately with
square root and quadratic dependences, respectively. 

\end{abstract}

\keywords{galaxies: active --- galaxies: blazars: individual
(3C~279) --- ultraviolet: galaxies --- ultraviolet: spectra ---
X-ray: galaxies}

\section{Introduction}

The blazar 3C~279 ($z$ = 0.538) is a strong emitter at all energies, and
one of the two brightest sources detected by EGRET in the $\gamma$-rays
(Hartman et al. 1992; Wehrle et al. 1998;  Mattox et al. 1997). Studies of
its UV and X-ray emission (Bonnell, Vestrand, \& Stacy 1993; Shrader et al.
1994; Koratkar et al. 1998; Makino et al. 1989) have detected remarkable 
variability on a range of time scales from days to years. 
Its UV spectrum exhibits strong, broad, high-ionization emission
lines (Netzer et al. 1994; Koratkar et al. 1998), albeit not so luminous
and broad as typically found in quasars, which makes this source
intermediate between BL Lac objects and highly polarized quasars.  The
spectral energy distribution of 3C~279 has two broad humps peaked at
far-infrared and MeV frequencies,
and identified, according to the general interpretation of blazar continua,
with synchrotron radiation and inverse Compton scattering, respectively, 
in a relativistic jet (Ulrich, Maraschi, \& Urry 1997). 

In the past few years, 3C~279 has been selected for simultaneous
radio-to-$\gamma$-ray monitoring. The coordination of space- and
ground-based observing facilities yielded the richest multiwavelength
variability data ever obtained for a blazar (Maraschi et al. 1994;
Hartman et al. 1996;  Wehrle et al. 1998).  The $\gamma$-ray emission, as
observed also for other blazars (PKS~1622--297, Mattox et al. 1997; 
PKS~0537--441, Hartman 1996; Mkn~421, Macomb et al. 1995, Gaidos et al. 
1996;  Mkn~501, Catanese et al. 1997, Aharonian et al. 1998; BL Lac,
Bloom et al. 1997), exhibited the largest amplitude variability (up to a
factor of $\sim$100) within the electromagnetic spectrum on different
time scales. However, these campaigns have not been able to determine the
nature of the photons upscattered to the highest energies through the
inverse Compton mechanism, and the origin of the $\gamma$-ray flares, a
major puzzle of blazar physics.

Different scenarios have been proposed (see also Ghisellini \& Maraschi
1996): the seed photons for the inverse Compton scattering can belong to
an optical-UV radiation field internal to the jet (synchrotron photons;
e.g., Maraschi, Ghisellini, \& Celotti 1992)  or external (accretion disk
or broad emission line region; Dermer \& Schlickeiser 1993; Sikora,
Begelman, \& Rees 1994).  The latter might be viable in sources for which
isotropic thermal and/or line emission is important, as compared with the
beamed, non-thermal radiation.  The strong emission lines observed in
3C~279 might provide the inverse Compton seed photons, which raises
however a further question about the powering mechanism of the line
emitting gas.  The small variability of the Ly$\alpha$ emission line of
3C~279, at least on long time scales, points to a modestly variable
ionizing source (Koratkar et al. 1998). This could be more likely a
source of thermal origin, such as an inner disk, rather than the beamed
radiation produced within the relativistic jet.  In fact, if a jet
illuminates the line emitting gas clouds, its beamed highly variable
emission would produce rapid and large changes in the line flux, although
only over a narrow velocity range.  These could be responsible for the
fast $\gamma$-ray continuum variations (Ghisellini \& Madau 1996; Wehrle
et al. 1998), however, short time scale (days or hours) line variability
has been poorly studied, and never observed so far, in 3C~279. 

The UV continuum of 3C~279 usually has a power-law shape and is rather
steep, revealing its non-thermal origin.  It is due to the radiation
losses of the highest energy relativistic electrons injected in the jet.
However, if the inverse Compton scattered photons are radiated from a
disk, or from a broad line region powered by a disk, this component ought
to be observed, at least when the synchrotron emission is quiescent. 
Therefore, 3C~279 in a low state is a good candidate for studying the
underlying thermal emission component.

In this paper, we concentrate on the IUE monitoring of 3C~279 conducted
during the EGRET observations in Phase 2 (December 1992-January 1993), when
the source was in a low multiwavelength emission state, and Phase 3
(December 1993-January 1994).  Some of the data were shown in Maraschi et
al. (1994), Koratkar et al.  (1998), and Wehrle et al. (1998). We present
here the results of an improved, systematic analysis of the IUE data and
compare them with X-ray results from the coordinated ROSAT PSPC and ASCA
observations conducted during the EGRET pointings in Phases 2 and 3,
respectively.  The X-ray data have been previously presented by Sambruna
(1997, ROSAT) and Kubo et al. (1998, ASCA).  Our independent re-analysis of
these data yielded completely consistent results, therefore we refer the
reader to those papers for a detailed description of the reduction and
analysis. 

In \S~2 we present the IUE observations and data analysis, in \S~3 we
describe the UV spectral shape, also in relation to the simultaneous optical
and X-ray data, and the light curves, and in \S~4 we discuss these results
and review the correlation of historical UV light curves of 3C~279 with the
X-ray and $\gamma$-ray light curves.

\section{IUE Data Acquisition, Reduction and Analysis}

Low dispersion spectra of 3C~279 were taken, in both IUE wavelength 
ranges,
1230-1950 \AA\ and 2000-3000 \AA, from 1992 December 19 to 1993 January 10,
and only at short wavelengths on 1993 December 24 and 25 (see Journal of
Observations in Table~1).  The source was in a very faint state, so that the
signal was weak relative to the background. Inspection of the
line-by-line LWP spectral images reveals contamination from solar scattered
light longward of $\sim$2750 \AA\ (Caplinger 1995, and references therein).
The NEWSIPS method for the spectral signal extraction, which has been
adopted for the implementation of the IUE Final Archive, was modified to
handle noisy, background-contaminated, low signal-to-noise
ratio
spectra.  The extraction routine uses a default profile (point source in the
present case)  instead of the empirical cross-dispersion profile (Imhoff
1996).  This avoids erroneous flux values for uncontaminated regions of the
spectrum, and also minimizes the contribution of the spatially extended
solar flux.  Therefore, we used for our analysis the NEWSIPS extracted
spectra. This extraction routine includes a camera head amplifier
temperature correction and a time-dependent sensitivity degradation
correction (see Nichols \& Linsky 1996, and Garhart et al. 1997,
where spectral flux calibration is
also discussed).  A recent improvement of the sensitivity degradation
correction has been considered for the SWP spectra taken after January 1993
(Garhart et al. 1997; Garhart 1998;  Imhoff 1997). For this reason, we
re-analyzed also the two SWP spectra (53261 and 56635) of 3C~279
taken after
the campaigns discussed in this paper (see Table~1).  Cosmic rays in
individual spectra were removed, and regions affected by camera artifacts
were discarded (Crenshaw, Bruegman, \& Norman 1990).  We estimated the flux
levels in the SWP and LWP spectra by averaging the signal in 200 \AA-wide
bands around the effective wavelengths of 1750 \AA\ and 2600 \AA\ for SWP
and LWP spectra, respectively. These fluxes, along with their 1-$\sigma$
uncertainties estimated as in Falomo et al. (1993), are reported in Table~1
and, after correction for Galactic reddening (see below), in Figure 1b (LWP
only). 

We have also compared our UV fluxes with those reported in Maraschi et al. 
(1994), using the same wavelength intervals and extinction value.  While the 
SWP signal is consistent within the errors, the LWP signal
reported earlier is almost a factor of 2 higher than presently found.  We 
ascribe
this discrepancy to the different spectral extraction.  The method used by
Maraschi et al. is likely to have significantly underestimated the solar light
contamination, resulting in higher LWP source fluxes.

The expected interstellar extinction in the direction of 3C~279 due to
Galactic neutral hydrogen is $A_V \simeq 0.13$ mag, corresponding to a
column density N$_{HI} = 2.22 \times 10^{20}$ cm$^{-2}$ (Elvis, Lockman, \&
Wilkes 1989), assuming a gas-to-dust ratio N$_{HI}/E_{B-V}$ = 5.2$\times
10^{21}$ cm$^{-2}$ mag$^{-1}$ (Shull \& Van Steenberg 1985) and a
total-to-selective extinction ratio $A_V/E_{B-V} = 3.1$ (Rieke \& Lebofsky
1985). Pairs of SWP and LWP spectra taken close together in time were
de-reddened with the extinction curve of Cardelli, Clayton, \& Mathis
(1989), and fitted to simple power-law models through an iterative,
chi-squared minimization routine, by avoiding wavelengths dominated by
Ly$\alpha$ emission at the redshift of the source (1850-1880 \AA), the
noisy regions shortward of 1230 \AA\ and between 2000 and 2400 \AA, and the
wavelength interval longward of 2700 \AA, which is contaminated by solar
scattered light.  The best-fit energy indices $\alpha_{UV}$ ($f_\nu \propto
\nu^{-\alpha}$) are reported in Table~2, as are the reduced $\chi^2$ values
for each fit ($\chi^2_\nu$). A photometric uncertainty of 1.25\% has been
added in quadrature to the statistical error associated with each fitted
flux (see Edelson et al.  1992). 

The $\chi^2_\nu$ values for the power-law fits are generally not
satisfactory.  This is probably due to the fact that the errors
associated with IUE spectral fluxes might be underestimated (see
Urry et al.  1993).  However, we tried instead fits with a
black-body model to the merged spectra in the same wavelength
interval used for the power-law fits. The results are reported in
Table~2.  The $\chi^2_\nu$ values are still not satisfactory, though
systematically smaller than those associated with the power-law
model fits.  No fit to single SWP or LWP spectra was attempted
because the former are too noisy for a reliable measurement of the
spectral index, and for the latter the wavelength baseline is
too narrow.

For comparison with the NEWSIPS method, we also retrieved and
analyzed the spectra of 3C~279 of January 1993 extracted with the
recently developed INES routine and available in the IUE archive
at Vilspa (Loiseau \& Schartel 1998).  Power-laws have been fitted
to the pairs of simultaneous SWP and LWP spectra after excluding
the above wavelength intervals and de-reddening the data. The
results are reported in Table~2. 

The $\chi^2_\nu$ values obtained by fitting power-laws to INES
extracted spectra are smaller than those of NEWSIPS spectra.  We
note however that spectral flux distributions derived from the
INES extraction are systematically higher than the NEWSIPS, with
differences in the SWP spectra being marginally significant, and
those in LWP spectra being much larger (up to 100\%).  This
results in steeper spectral indices for fitted power-laws on
average ($\langle \alpha_{UV} \rangle \simeq 1.5 \pm 0.2$, see
Table~2), although still flatter than expected, based on the
emission state (see \S 3.1 and cfr. with Table~3).  The fact that
the NEWSIPS LWP flux smoothly connects with the simultaneous
optical data (Fig. 4) makes us suspect that the INES routine
removes the contribution of the scattered light only partially,
thus yielding significantly higher LWP fluxes than NEWSIPS does. 
Therefore, we prefer the NEWSIPS extraction. We also
stress that at the present flux levels, slight differences in the
background estimate either in the SWP or LWP camera can lead to
dramatic differences in the extracted signal.

\section{Results}

\subsection{The Ultraviolet Spectral Shape}

The power-law fit to the quasi-simultaneous pairs of NEWSIPS extracted
SWP and LWP spectra yields a much flatter energy index ($\langle
\alpha_{UV} \rangle \simeq 1.0 \pm 0.2$) than generally measured in the
1230-3000 \AA\ range for this source ($\langle \alpha_{UV} \rangle \simeq
1.5$; Webb et al.  1990;  Edelson et al.  1992; Bonnell et al.  1994;
Shrader et al.  1994). We investigated whether this can be ascribed to
the more limited wavelength range used here (1230-2700 \AA\ instead of
1230-3000 \AA), or to a mismatch between the SWP and LWP spectra due to
the NEWSIPS extraction routine or the calibration. To this purpose, we
retrieved all the pairs of quasi-simultaneous (within $\sim$1 day) 
NEWSIPS SWP and LWP spectra of 3C~279, fitted them jointly in the
1230-2700 \AA\
interval, and compared our results (see Table~3)  with those of the above
authors (who used different extraction methods), finding a very good
agreement. In the last line of Table~3 are reported the results of the
fit to the joined simultaneous SWP and LWP spectra taken in the last
IUE campaign of 3C~279 (Jan-Feb 1996, Wehrle et al.  1998), during which
the UV flux was 3 times higher, which indicate a spectral slope steeper
than those reported in the previous studies.  We conclude the difference in
spectral shape is real, not an artifact of the analysis.
(Note that the extinction correction adopted here, $A_V
= 0.13$, is the same as or higher than in previous analyses of the UV
data.  Therefore, to rule out de-reddening as a cause of spurious
spectral flatness, we report in Table~3 the best-fit spectral indices for
both observed and de-reddened spectra.) 
We also checked that no saturated spectra had been taken prior to each of our
SWP exposures, which could have caused spurious residual flux adding up
to the signal from 3C~279, and verified that the background in a
region of the image very close to the object spectrum was not
underestimated during spectral extraction. Although the signal-to-noise
ratio of these spectra is rather low, we find no instrumental explanation
for the unusual flatness of the spectrum and so we conclude it has a
physical origin.  

In Figure 2 we show the available UV data (both from IUE and the HST Faint
Object Spectrograph, FOS) of 3C~279, corrected for Galactic absorption and
averaged, to increase the signal-to-noise ratio, according to flux level
at the different epochs of observation (see caption). Superposed are the
best fit power-law curves for each state. These fits show that the
power-law slope flattens in the very low state of January 1993, as opposed
to a general spectral steepening with decreasing flux seen at the other
epochs, when the emission state was higher. The extremely low spectrum of
January 1995 is also shown: it is similarly flat.  Figure 3 shows the flux
versus the spectral index for all the spectra in Figure 2.  We have estimated
the UV flux in a way independent of the fitted spectral index, to avoid
introducing spurious correlation between the two quantities.  For IUE
spectra, the UV flux is the geometrical mean of the SWP signal at 1750
\AA\ and the LWP signal at 2600 \AA, therefore it refers to an effective
wavelength of $\sim$2130 \AA.  Uncertainties have been derived as in
Edelson et al.  (1992).  For HST spectra, the flux at 2130 \AA\ has been
directly measured.  The figure illustrates clearly the spectral steepening
accompanying the flux decrease, and, at a flux of $\sim$1 mJy, the
reversal of this trend. 

\subsection{Comparison with the Optical Spectral Shape}

To better characterize the spectral behavior of 3C~279 and elucidate the
cause of the anomalous UV spectral flatness, we compared the UV data of
January 1993 with optical BVRI photometry performed quasi-simultaneously
(within $\sim$12 hours) with the IUE observations (Grandi et al. 1996).
In Table~2 we report the indices, along with $\chi^2_\nu$ values, of
power-laws fitted to the optical data only, optical and LWP, and optical,
LWP and SWP. (For these fits, the UV data have been grouped in $\sim$100
\AA\ bins.)  These parameters indicate that (1) the slopes of optical
spectra are significantly steeper than those of the simultaneous UV
spectra; (2) the indices of optical spectra are very similar to those of
combined optical and LWP spectra, while (3) they are steeper than those
of combined optical, LWP and SWP spectra.  This suggests that the LWP
spectra are consistent with the extrapolation of the optical to UV
wavelengths, but that shortward of $\sim$2000 \AA\ a remarkable spectral
flattening occurs. 

The optical-UV spectral shape therefore
suggests the presence of an emission component appearing at
$\sim$2000 \AA\ superposed on that producing the optical and
near-UV flux. One obvious candidate is thermal radiation from an
inner accretion disk.  Fitting the 1230-2700 \AA\ flux
distribution to a black-body model yields fit parameters for the
four spectra consistent at the 2-$\sigma$ level (Table~2), and an
average temperature of $\sim$20000~K.  Note that the black-body
model fits have slightly smaller $\chi^2_\nu$ values than
the power-law fits. However the very small difference, and
the fact that $\chi^2_\nu$ values might be too high due to an
underestimate of the IUE flux errors, do not allow us to select
the correct model based only on the $\chi^2$.  As a consistency
check, black-body fits to the higher state average spectra
reported in Figure 2 all give worse $\chi^2_\nu$ values than power-law
fits.

\subsection{Comparison with the X-ray Spectral Shape}

We also explored whether the UV spectral flattening observed in January
1993 could have affected the simultaneously measured soft X-ray spectrum. 
The energy index for the ROSAT spectra in December 1992-January 1993 has
an average $\langle \alpha_X \rangle = 0.85 \pm 0.05$ (0.1-2.4 keV), with
very small variability (Sambruna 1997).  ASCA observations one year later
yield a significantly harder spectrum ($\langle \alpha_X \rangle = 0.65
\pm 0.07$ in the range 2-10 keV, with a $\chi^2_\nu = 0.57$; Kubo et al.
1998) than the ROSAT one, but spectral variability between the two epochs
is hard to assess, because of the difference in the energy ranges of the
ROSAT and ASCA detectors, which are only partially overlapping.  If
temporal variability could be excluded, the observed difference would
suggest that the spectrum hardens with energy in the 0.1-10 keV range. 
This is supported by the fact that the slope of the ASCA spectrum in the
0.7-10 keV energy range is $0.75 \pm 0.03$, namely softer than between 2
and 10 keV. The average spectral change between the UV and X-ray
simultaneous spectra in January 1993 is $\Delta \alpha \sim 0.2$. We also
retrieved and analyzed the so far unpublished ASCA spectra of 3C~279
taken in December 1994-January 1995, obtaining a fitted spectral index
$\alpha_X = 0.64 \pm 0.04$ ($\chi^2_\nu \simeq 1$) in the 2-10 keV range.
Including the softest energies in the fit (0.7-2 keV) does not result in
a significantly different slope (the parameter N$_{HI}$ was fixed at the
Galactic value). 

In the first four panels of Figure 4 we report the de-absorbed
simultaneous optical-to-X-ray energy distributions of 3C~279 in January
1993, along with the black-body fit to the average IUE spectrum (fitted
temperature $T = 20000 \pm 1000$ K, $\chi^2_\nu = 0.8$). In the fifth
panel we show a brighter state, for comparison, from the IUE-SWP and ASCA
spectra in December 1993, and the average IUE-LWP spectrum of January
1996 (see caption to Fig. 2), which, albeit non-simultaneous with the SWP
data, pertains to an emission state of similar level.  No optical data at
this epoch are available. The bright UV spectrum follows a power-law and
any (flat)  thermal component, if present at the level seen one year
earlier, is not detectable. 

\subsection{Light Curves and Correlations}

The UV flux at 1750 \AA\ does not vary significantly during the first
campaign (Jan 93), but increases by almost a factor of 3 the next year
(Table~1). There is also a variation of $\sim$45\% between the latter two
observations in December 1993.  At longer UV wavelengths (2600 \AA), for
which sampling is available only during the first campaign, the flux
varies day-to-day by 20-30\%, if one excludes the 3-$\sigma$ upper limit
on 1992 December 27 (Fig.  1b).  Including the upper limit, the maximum
variation is more than 50\% in a week.  For comparison, the optical flux
steadily increases during the monitoring by a factor 2.5 or more (Fig. 
1a).  The ROSAT light curve shows a $\sim$13\% amplitude flare of 2 days
duration at the beginning of the campaign and a symmetric, larger (20\%
flux change from low to high state), slower ($\simlt$10 days including
increase to maximum and decrease to initial state), and better sampled
flare toward the end (Fig. 1c). 
An increase of $\sim$50\% is observed between the average ROSAT flux at 1
keV ($\sim 1~\mu$Jy, Sambruna 1997) and that recorded by ASCA one year
later ($\sim 1.5~\mu$Jy, limiting the comparison to the soft energy
band of ASCA, 0.5-2 keV).  

We used the Discrete Correlation Function (DCF, Edelson \& Krolik
1988) to study the correlation between X-ray emission and spectral slope
during the ROSAT campaign (Fig. 2c and Table 2 of Sambruna 1997).  The DCF
amplitude
curve (Fig. 5) has a minimum at a positive time-lag of $\sim$2-3 days,
suggesting that the flux increase leads the flattening of the spectral
shape (hard lag). 

\section{Discussion}

\subsection{A Possible Thermal Origin of the Ultraviolet Radiation in
the Low State}

We have observed 3C~279 at UV wavelengths at two epochs separated by
one year, simultaneously with X- and $\gamma$-ray observations. 
During the first epoch (Dec 92-Jan 93), the source was in one of its
dimmest UV states ever. (The absolutely lowest state was reached in
January 1995, see Table~1 and Fig. 2.) At the second epoch (Dec 93),
the UV flux was a factor of 3 higher.

The archival IUE observations of 3C~279 suggest anticorrelation of the
UV flux and spectral slope (flatter spectrum for brighter flux)  at
high or moderate emission levels, but the opposite behavior --
positive correlation -- at lower emission states.  Indeed, at the
lowest fluxes, the IUE spectral index is much flatter, $\langle
\alpha_{UV} \rangle \sim 1$, than ever before seen.  The high
signal-to-noise ratio HST-FOS data support this reversal of trend, in
that the 1992 spectrum, which corresponds to a lower state than that
seen with the FOS in 1996, has also a slightly flatter slope. The flat
slope of the IUE spectra can be fitted with a black-body model, with
temperature $T \sim 20000$ K, but, because of the faintness of the
source, it is impossible, based only on the goodness of the fits, to
rule out a power-law model in favor of a black-body or other thermal
(curved)  model. 

The optical spectrum is smooth and likely produced by synchrotron
radiation (Grandi et al. 1996), with a slope $\alpha_{opt} \sim$
1.8-2.0. Thus, significant spectral flattening occurs in the UV range. 
If the UV spectrum were also due to the synchrotron mechanism it would
imply a relativistic electron distribution that flattens sharply at
higher frequencies whereas the electron distribution generally
steepens progressively at higher frequencies due to radiative losses. 
Moreover, the optical flux varies with much higher amplitude than the
UV (Fig. 1), which is also opposite to what is expected in a pure
synchrotron model. It is possible that the flattening of the UV
spectrum observed in the low state is due to a contribution of inverse
Compton scattering extending to the UV wavelengths.  However, the
fitted temperature of the black-body model and the luminosity derived
from the fitted normalization (see below) make it plausible
that the low state UV spectrum is produced by thermal
emission from an accretion disk with a temperature $T \sim 20000$ K. 
At soft X-ray energies the contribution of the black-body component
falls off rapidly, and at 0.1 keV is negligible ($\sim 10^{-20}$ of
the total flux), but one cannot exclude a significant contribution in
the X-rays under the hypothesis of a more complex model, like a
multi-temperature disk.

The spectral flattening observed in the 1230-2700 \AA\ range at very
low UV continuum levels, an uncommon feature in blazars, is typical
of radio-quiet quasars.  Quite plausibly it could be the signature of
a thermal, quasi-isotropic component, perhaps a disk, which is
normally swamped by the non-thermal, highly variable, beamed
continuum of 3C~279.  This ``Seyfert-like" component would be visible
in the UV only in very low states, namely when the contribution of
the synchrotron radiation is less important.  (The quality of the
present UV data does not allow us to disentangle the two
contributions by means of composite fits.)  This could also explain
the smaller variability observed in the UV than in optical light,
under the assumption that the disk emission is not variable on time
scales less than a few years. The
observation of a still weaker flux in January 1995 at the shorter UV
wavelengths (Fig. 2) suggests some yearly variability of this thermal
component, albeit modest.  Observation of wavelength dependent
polarization in the UV could confirm the thermal interpretation if it
is markedly different in the low state.

Assuming that a single black-body is a good approximation for the UV
emission, and using the fit parameters obtained for the average IUE
spectrum, at $z = 0.538$, with $H_0 = 65$ km s$^{-1}$ Mpc$^{-1}$ and $q_0
= 0.5$, the fitted normalization of the black-body corresponds to a
luminosity $L_{UV} \sim 2 \times 10^{45}$ erg s$^{-1}$ (if isotropic) and
to a linear size of the emitting region of $\sim$1 light day, consistent
with the range of UV luminosities of typical quasars (Elvis et al. 1994)
and with the inner dimensions of an accretion disk in an active galactic
nucleus (Rees 1984), respectively. Theoretically, the presence of an
accretion disk has been invoked as the hidden power source of the broad
H$\alpha$ emission line in BL Lac (Corbett et al. 1996). The absence of
intensity variations in the strong, broad Ly$\alpha$ emission line of
3C~279, as opposed to the high amplitude variability of the UV continuum,
also argues for the presence of an accretion disk powering the broad line
region (Koratkar et al. 1998).  If this were the case, the UV
photoionizing flux provided by our fitted black-body component ($\sim 1.4
\times 10^{-12}$ erg s$^{-1}$ cm$^{-2}$) would account for the observed
Ly$\alpha$ line intensity ($\sim 6.5 \times 10^{-14}$ erg s$^{-1}$
cm$^{-2}$) assuming values normally expected in AGNs for the covering
factor of the broad line clouds ($\sim$5\%, e.g., Netzer 1990). 

The fitted temperature of the black-body, 20000~K, implies an emission
peak at rest frame wavelength $\sim$1500 \AA, in the range between the
so-called ``blue bump", seen in quasars and interpreted as accretion disk
emission, and the soft X-ray excess, exhibited by several Seyfert
galaxies and generally identified as the high energy tail of the blue
bump itself (see Kolman et al. 1993; Czerny \& Elvis 1987; reviews by
Bregman 1990 and 1994).  A suggestion of an optical excess was found in a
few blazars (Brown et al. 1989) and a blue bump has been clearly observed
in at least one of them, 3C~345 (Bregman et al.  1986), in the radio-loud
quasar 3C~273 (Shields 1975;  Ulrich 1981), and in the radiogalaxy 3C~120
(Maraschi et al. 1991).  For 3C~273 and for another radio-loud quasar,
B2~1028+313, a soft X-ray excess has been detected (Masnou et al. 1992;
Grandi et al. 1997;  Haardt et al. 1998).

\subsection{Multiwavelength Light Curves and Spectra}

Multiwavelength variability (Maraschi et al. 1994; Hartman et al. 1996; 
Wehrle et al. 1998; Maraschi 1998) suggests that the broad-band spectrum
of 3C~279 at radio-to-UV wavelengths and at X- and $\gamma$-ray energies
is produced by synchrotron radiation and inverse Compton scattering,
respectively.  This is supported by the soft, power-law optical and UV
continuum and by the flat spectra observed at medium and hard X-rays
(Makino et al.  1989;  Hartman et al.  1996;  Kubo et al.  1998; Lawson
\& McHardy 1998).  The spectral variability seen above $\sim$2 keV is
small, consistent with the fact that the X-rays are produced through
inverse Compton scattering of relativistic electrons with small Lorentz
factor, whose energy distribution slope is not expected to vary on short
time scales.  In fact, our analysis of the ASCA spectra of December
1994-January 1995 yields an energy index identical to or not
significantly different from those found at other epochs.  The
anticorrelation between UV flux and spectral index variations, predicted
by models based on radiative cooling (Celotti, Maraschi, \& Treves 1991),
is instead consistent with a distribution of electrons with high Lorentz
factor, which have more rapid radiative losses. 

The soft X-ray spectrum appears more variable in the long term and it is
softer for fainter flux (Schartel et al.  1996;  Sambruna 1997), which
might indicate that other radiation processes besides inverse Compton 
contribute in this spectral region, like, for instance, the high energy
tail of the synchrotron component.  Another possible candidate is thermal
radiation. Although there is no sign in joint RXTE and ROSAT HRI data of
a soft X-ray excess of thermal origin in this source (Lawson \& McHardy
1998), the 0.1-2 keV spectrum measured by ROSAT in December 1992-January
1993 is steeper than that measured by ASCA in the 2-10 keV band in
December 1993, which suggests the presence of an extra component at the
lower energies, assuming no X-ray spectral variability between those two
epochs. Moreover, in December 1993 the spectral slope in the range 0.7-10
keV was somewhat steeper than in 2-10 keV, which again points to the
above suggestion.

Note that the soft X-ray flux variations precede those of the spectral
index (Fig. 5), which is the contrary of what has been observed in other
blazars, like Mkn~421 and PKS~2155--304, both in X-rays (Takahashi et al.
1996; Sembay et al. 1993; Treves et al. 1999; Chiappetti et al. 1999) 
and UV (Pian et al. 1997). In those sources the UV and X-ray emission is
dominated by synchrotron radiation, and the variations at harder energies
lead those at softer energies, which determines a lag of flux with
respect to spectral index.  In 3C~279 the non-thermal X-ray emission
likely has a different origin (inverse Compton instead of synchrotron)
and in addition it might be diluted by thermal radiation at the softer
energies. 

From the numerous multiwavelength monitorings of 3C~279 we collected the
X- and $\gamma$-ray data simultaneous with IUE measurements and report in
Figure 6 the UV flux at 2000 \AA\ versus the 2 keV X-ray flux (Fig. 6a)
and versus the 400 MeV $\gamma$-ray flux (Fig. 6b).  When simultaneous
pairs of SWP and LWP spectra were available, the flux at 2000 \AA\ was
obtained from the joint power-law fit of the two spectra (see Tables 2
and 3);  otherwise, it was obtained by extrapolating to this wavelength
the available flux (Table~1)  using the typical spectral index for the
corresponding emission level (see Fig. 2; for January 1995, the average
spectral index of January 1993 has been used). In 1991, no strictly
simultaneous $\gamma$-ray and UV data are available, therefore we
associated with the EGRET data of June 1991 the IUE measurements of June
1989, based on the similarity of the optical emission state in June 1989
and June 1991 (see Hartman et al. 1996).  Both X- and $\gamma$-ray
emission appears to be generally well correlated with the UV emission,
although the X-ray flux varies somewhat less than the UV, very roughly as
the square root of it, as found by a linear regression test, while the
$\gamma$-ray flux varies roughly as the square of the UV flux. The
UV-to-X-ray flux correlated variability between January 1993 and December
1993 is in agreement with this historical trend, in that the X-ray flux
increased only by 50\%, to be compared to the factor of 3 variation seen
in the UV.  

On long time scales, these correlations support both the scenario in
which the inverse Compton seed photons are the synchrotron photons
themselves (synchrotron self-Compton) and that in which the photons
upscattered to the highest energies are provided by a source external
to the jet, like radiation coming directly from an accretion disk, or
reprocessed in a broad line region (external Compton).  In the former
case, the $\gamma$-rays are expected to vary more than the
infrared-to-UV flux, due to the non-linearity of the synchrotron
self-Compton mechanism (Ghisellini \& Maraschi 1996).  In the latter
case a linear correlation between the optical-UV and the $\gamma$-ray
flux variations would be expected, but higher amplitude $\gamma$-ray
variations could be reconciled with this prediction, provided the bulk
Lorentz factor of the relativistic plasma varies between the
observation epochs (Hartman et al.  1996).  Neither scenario can be
favored against the other, the relative importance of the two
depending on the bulk Lorentz factor (Ghisellini \& Maraschi 1996).
The largest quasi-simultaneously measured UV and $\gamma$-ray fluxes
of 3C~279 present an inverted correlation.  Although this might be
simply due to the non strict simultaneity of the data or to a
difference in the physical parameters at the two epochs, like e.g.,
magnetic field and maximum electron energy (the sampling time is
longer than the typical multiwavelength variability time scales), it
does not exclude the possibility that, at different epochs, the
$\gamma$-rays are produced through inverse Compton radiation off
optical-UV photons of different origin.

On the shortest time scales (one day or less) the observed variability
(the changes in the $\gamma$-rays are possibly more than quadratic with
respect to UV, Wehrle et al. 1998) is definitely incompatible with
external Compton scattering (changes in the bulk Lorentz factor are
unlikely to occur on day, or shorter, time scales), and could pose
difficulties also for the synchrotron self-Compton model.  An
alternative is the ``mirror" model (Ghisellini \& Madau 1996), in which
the target photons for the inverse Compton scattering consist of emission
line photons produced in a small number of broad line region clouds
illuminated by the relativistic jet.  The rapid variations caused by the
jet on the line emission can result, in the comoving frame of the jet, in
more than quadratic variations in the $\gamma$-ray flux radiated via
the inverse Compton mechanism. 

Our proposal of a thermal component underlying the UV continuum of 3C~279 is
critical for both ranges of time scales.  The UV luminosity we find for the
putative accretion disk responsible for this component is consistent with
the estimates provided by Dermer and Schlickeiser (1993)  and Sikora et al. 
(1994), who interpret the high energy spectra of 3C~279 in terms of an
inverse Compton scattered isotropic field radiated from a central source, an
accretion disk or the broad line region. We recall that the estimated UV
luminosity of the accretion disk makes it a reasonable powering source of
the Ly$\alpha$ emission line.  This line indicates the presence of broad
line clouds, some of which could interact with the jet, responsible of the
synchrotron continuum, and give rise to the mirror effect.  Rapid
fluctuations in subcomponents of the Ly$\alpha$ line profile would be
expected in this case.  Therefore, sensitive UV observations with high
spectral resolution would be important for discriminating the active
processes.  The presence of a thermal component needs to be confirmed by
more sensitive UV observations, as well as by more hard X-ray and
$\gamma$-ray data, during a low state of 3C~279.

\acknowledgements 

We are grateful to C. Cacciari, J. Caplinger, D. De Martino, and N. Loiseau
for their assistance with IUE observations and data reduction, and to the IUE
staff of Vilspa and GSFC, in particular to W. Wamsteker and Y. Kondo, for
their support of this project.  We thank R. Scarpa for useful comments to the
paper. This work was supported by NASA grants NAG5-2154, NAG5-2538, and
NAG5-3138.


\clearpage



\figcaption{Multiwavelength light curves of 3C~279 in December
1992-January 1993:  (a) dereddened optical R-band fluxes (from Grandi
et al.  1996);  (b)  dereddened IUE fluxes at 2600 \AA;  (c)
de-absorbed fluxes at 1 keV for Galactic $N_{HI} = 2.22 \times
10^{20}$ cm${-2}$ (from Sambruna 1997). Error bars represent
1-$\sigma$ uncertainties.}

\figcaption{De-reddened UV spectra of 3C~279 binned in $\sim$100
\AA-wide intervals.  Like for other synchrotron sources, the spectra
steepen with decreasing intensity. However, at the lowest intensity,
the spectra are again flat, possibly due to the presence of steady
accretion disk emission.  Error bars are 1-$\sigma$ uncertainties. 
Open symbols represent average IUE spectra for various typical flux
levels (see Table~3): the ``high" state spectrum (Jul 1988, circles) 
was obtained by co-addition, weighted with exposure time, of spectra
SWP 33864, 33865 for the 1230-2000 \AA\ range and LWP 13566, 13567 for
the range 2000-3000 \AA; the ``medium" state (Jan 1989-May 1992,
squares) is the co-addition of SWP 35443, 36420, 42132, 40489, 44806
and LWP 14933, 15677, 20891, 19492, 23207; the ``low" state (Dec 1993
and Jan 1996, triangles) has been obtained by co-adding SWP 49681,
49686, 56635 and LWP 31882, 31906, 31908, 31914; the ``very low" 
state (Jan 1993, diamonds) is the co-addition of SWP 46649, 46653,
46657, 46662 and LWP 24652, 24656, 24661, 24665; the stars represent
the ``extremely low" state observed in January 1995 (SWP 53261). 
Filled symbols represent HST-FOS spectra in April 1992 (squares) and
January 1996 (circles).  Superimposed on each representative spectrum
is the power-law curve which best fits the data in the interval
1230-2700 \AA. Spectral indices are $\alpha = 1.47 \pm 0.02$ (Jul
1988), $\alpha = 1.70 \pm 0.03$ (Jan 1989-May 1992), $\alpha = 1.92
\pm 0.07$ (Dec 1993 and Jan 1996), $\alpha = 0.89 \pm 0.15$ (Jan
1993), $\alpha = 1.65 \pm 0.13$ (HST-FOS 1992), $\alpha = 2.25 \pm
0.04$ (HST-FOS 1996).}

\figcaption{De-reddened UV flux at 2000 \AA\ versus energy index at
the various epochs of IUE and HST observations of 3C~279. The upper
branch shows the usual anti-correlation of spectral index and
intensity for bright synchrotron sources, while the lower branch
suggests the increasing importance of a flat accretion disk component
at the lowest UV fluxes.}

\figcaption{Simultaneous de-absorbed optical-to-X-ray spectral energy
distributions of 3C~279 in a faint state, at the four epochs of IUE
SWP and LWP observations in January 1993 (see Table~2), and
UV-to-X-ray spectral energy distribution in a brighter state, in
December 1993 (only the SWP fluxes are simultaneous with the ASCA
data, the LWP fluxes are from January 1996, see text).  The dotted
curve shown in each panel represents the black-body model which
best-fits the average UV spectrum of January 1993 ($T = 20000 \pm
1000$ K). The possible disk component is clearly visible in the first
four spectra, but disappears when the synchrotron radiation rises. 
The UV spectral fluxes are binned in $\sim$100 \AA-wide intervals and
the 1-$\sigma$ uncertainties are estimated as in Falomo et al.
(1993).  The best fit power-laws to the ROSAT spectra (Sambruna 1997)
and ASCA spectrum (0.7-10 keV, this paper) are shown along with their
90\% confidence ranges.  The optical data are from Grandi et al.
(1996).}

\figcaption{Amplitude of the Discrete Correlation Function (Edelson \&
Krolik 1988) between the ROSAT count rates and spectral slopes. The
positive time lag corresponds to spectral index changes trailing those of
the emission.}

\figcaption{Historical de-absorbed UV fluxes versus simultaneous (a)
X-ray and (b) $\gamma$-ray fluxes of 3C~279, on a logarithmic scale. 
Both panels show clear correlation, with X-rays varying slowly with
respect to the UV and $\gamma$-rays varying with higher amplitude than
the UV. X- and $\gamma$-ray points, except the ASCA data of January
1995 which have been analyzed by us, are from Maraschi et al. (1994),
Hartman et al.  (1996), Hartman (1996), Sambruna (1997), Kubo et al.
(1998), Wehrle et al. (1998).  The $\gamma$-ray data have been
converted to Jansky according to Thompson et al. (1996).}

\newpage

%
%
\begin{center}
\begin{tabular}{ccrcc}
\multicolumn{5}{c}{{\bf Table 1:} IUE Observations of 3C~279
in December 1992-January 1996}\\
\hline
\hline
IUE Image & \multicolumn{2}{c}{Observation Midpoint} & t$^a$ &
$F_\lambda^b$\\
& JD - 2,440,000 & \multicolumn{1}{c}{UT} &  & \\
\hline 
LWP 24540 &  8976.30897 & 1992 Dec 19.80897 & 180 & $0.51 \pm 0.07$ \\
LWP 24597 &  8983.14525 &          26.64525 & 130 & $< 0.3^c$ \\
LWP 24607 &  8984.15172 &          27.65172 & 115 & $0.51 \pm 0.15$ \\
LWP 24616 &  8985.98721 &          29.48721 & 156 & $0.33 \pm 0.08$ \\
LWP 24640 &  8988.27792 &          31.77792 & 164 & $0.45 \pm 0.07$ \\
LWP 24652 &  8990.28179 & 1993 Jan 02.78179 & 360 & $0.68 \pm 0.08$ \\
SWP 46649 &  8990.48576 &          02.98576 & 220 & $1.24 \pm 0.15$ \\
SWP 46653 &  8991.05085 &          03.55085 & 600 & $1.08 \pm 0.10$ \\
LWP 24656 &  8991.38942 &          03.88942 & 360 & $0.58 \pm 0.06$ \\
SWP 46657 &  8992.05343 &          04.55343 & 660 & $0.82 \pm 0.11$ \\
LWP 24661 &  8992.41147 &          04.91147 & 360 & $0.57 \pm 0.07$ \\
SWP 46662 &  8993.05675 &          05.55675 & 660 & $1.03 \pm 0.15$ \\
LWP 24665 &  8993.43048 &          05.93048 & 315 & $0.56 \pm 0.08$ \\
LWP 24699 &  8998.23747 &          10.73747 & 150 & $0.56 \pm 0.07$ \\
SWP 49681 &  9346.15726 &      Dec 24.65726 & 200 & $2.42 \pm 0.21$ \\
SWP 49686 &  9347.04185 &          25.54185 & 330 & $3.49 \pm 0.21$ \\
SWP 53261 &  9720.84844 & 1995 Jan 03.34844 & 280 & $0.67 \pm 0.17$ \\
SWP 56635 & 10107.94094 & 1996 Jan 25.44094 & 325 & $2.46 \pm 0.16$ \\
\hline  
\multicolumn{5}{l}{$^a$ Exposure time in minutes.}\\   
\multicolumn{5}{l}{$^b$ Observed average flux in the range 1650-1850   
\AA\ for SWP spectra and }\\
\multicolumn{5}{l}{ ~~ 2500-2700 \AA\ for LWP spectra.  Units are
$10^{-15}$ erg s$^{-1}$ cm$^{-2}$ \AA$^{-1}$.}\\
\multicolumn{5}{l}{$^c$ 3-$\sigma$ upper limit.}\\
\multicolumn{5}{l}{Note. Errors represent 1-$\sigma$ uncertainties.}\\
\end{tabular}
\end{center}
\newpage
%
%
\begin{center}
\begin{tabular}{lcccc}
\multicolumn{5}{c}{{\bf Table 2:} Fit Parameters of De-reddened Spectra$^a$}\\
\hline
\hline
UT$^b$ (Jan 1993) & 02.88378 & 03.72014 & 04.73245 & 05.74362 \\
\hline
$\alpha^c$ (1230-2700 \AA) &  $1.37 \pm 0.18$ & 
                              $0.75 \pm 0.13$ & 
                              $0.60 \pm 0.16$ & 
                              $1.17 \pm 0.16$ \\
$F^d$ (mJy) & $0.139 \pm 0.007$ & 
              $0.153 \pm 0.006$ &
              $0.145 \pm 0.008$ & 
              $0.147 \pm 0.007$ \\
$\chi^2_\nu$ ($N_{dof}$) & 1.55 (499) & 1.72 (513) & 
1.64 (513) & 2.07 (513) \\ 
$\alpha^e_{INES}$ (1230-2700 \AA) & $1.53 \pm 0.18$ & 
                              $1.51 \pm 0.13$ &
                              $1.90 \pm 0.15$ & 
                              $1.13 \pm 0.18$ \\
$F^f_{INES}$ (mJy) & $0.19 \pm 0.01$ &
                   $0.210 \pm 0.008$ &
                   $0.212 \pm 0.008$ & 
                   $0.17 \pm 0.01$ \\
$\chi^2_\nu$ ($N_{dof}$) & 1.11 (499) & 1.31 (513) & 1.11 (513) & 
1.41 (513) \\ 
$\alpha$ (1230-8000 \AA) & $1.52 \pm 0.04$ & 
                    $1.62 \pm 0.04$ & 
                    $1.64 \pm 0.04$ & 
                    $1.76 \pm 0.05$ \\ 
$\chi^2_\nu$ ($N_{dof}$) & 5.21 (11) & 3.76 (12) & 4.07 (12) & 2.61 (12) \\
$\alpha$ (2400-8000 \AA) & $1.82 \pm 0.06$ &
                    $1.91 \pm 0.06$ & 
                    $1.91 \pm 0.07$ & 
                    $2.06 \pm 0.09$ \\
$\chi^2_\nu$ ($N_{dof}$) & 0.94 (5) & 1.39 (5) & 2.07 (5) & 1.27 (5) \\
$\alpha$ (4400-8000 \AA) & $1.88 \pm 0.08$ & 
                    $1.80 \pm 0.09$ & 
                    $1.78 \pm 0.08$ & 
                    $1.96 \pm 0.11$ \\
$\chi^2_\nu$ ($N_{dof}$) & 1.32 (2) & 0.29 (2) & 0.38 (2) & 0.05 (2) \\
$T_{BB}^g$ (K) & $17000 \pm 800$ & $20000 \pm 1000$ & $22000 \pm 1400$ 
& $17800 \pm 850$ \\
$F^h_{BB}$ (mJy) & $0.16 \pm 0.08$ & $0.17 \pm 0.07$ & $0.15 \pm 0.09$ 
& $0.16 \pm 0.08$ \\
$\chi^2_\nu$ ($N_{dof}$) & 1.52 (499) & 1.67 (513) & 1.62 (513) & 
2.03 (513) \\
\hline 
\multicolumn{5}{l}{$^a$ The NEWSIPS extracted IUE spectra have been 
used, unless otherwise}\\
\multicolumn{5}{l}{ ~~ indicated. Reduced $\chi^2$ values, with 
relative number of degrees of freedom}\\
\multicolumn{5}{l}{ ~~ in parentheses, refer to the fit parameters 
preceding them.}\\
\multicolumn{5}{l}{$^b$ Observation midpoint of quasi-simultaneous
SWP and LWP spectra}\\
\multicolumn{5}{l}{ ~~ (see Table~1).}\\
\multicolumn{5}{l}{$^c$ Fitted power-law index ($F_\nu \propto
\nu^{-\alpha}$).}\\
\multicolumn{5}{l}{$^d$ Fitted power-law normalization at 2000 \AA.}\\
\multicolumn{5}{l}{$^e$ Power-law index fitted to INES extracted IUE 
spectra.}\\
\multicolumn{5}{l}{$^f$ Power-law normalization fitted to INES 
extracted IUE spectra at 2000 \AA.}\\
\multicolumn{5}{l}{$^g$ Fitted black-body temperature.}\\
\multicolumn{5}{l}{$^h$ Fitted black-body normalization at 2000 \AA:}\\
\multicolumn{5}{l}{ ~~  $F_{BB} = N
({\nu \over\nu_0})^3 [exp(h \nu/kT) - 1]^{-1}$,
$N = 2\pi h\nu_0^3\Sigma/c^2$, where}\\
\multicolumn{5}{l}{ ~~ $\Sigma$ = angular size of the emitting region 
in steradians,
and $\nu_0 = 1.5 \times 10^{15}$ Hz.}\\
\multicolumn{5}{l}{Note. Errors represent 1-$\sigma$ uncertainties.}\\
\end{tabular}
\end{center}
%
%
%
\begin{center}
\begin{tabular}{cccccccc}
\multicolumn{8}{c}{{\bf Table 3:} IUE Final Archive Merged Spectra of 
3C~279:}\\
\multicolumn{8}{c}{Power-Law Fit Parameters$^a$}\\
\hline
\hline
\multicolumn{2}{c}{Spectral Pair} & Date & $\alpha_{UV}$ &  
$\chi^2_\nu$ &  $\alpha_{UV,0}$ & $F_\nu^b$ & $\chi^2_\nu$ \\
 SWP  & LWP & & &  & & (mJy) & \\
\hline
33864 & 13566 & 1988 Jul 06 & $1.76 \pm 0.02$ & 1.52 &
$1.63 \pm 0.03$ & $3.69 \pm 0.05$ & 1.47 \\
33865 & 13567 & 1988 Jul 06 & $1.51 \pm 0.03$   & 1.89 &
$1.36 \pm 0.03$ & $3.91 \pm 0.06$ & 1.89 \\
35443 & 14933 & 1989 Jan 29 & $1.61 \pm 0.03$   & 1.28 &
$1.47 \pm 0.03$ & $1.72 \pm 0.03$ & 1.30 \\
36420 & 15677 & 1989 Jun 09 & $1.85 \pm 0.02$   & 1.80 &
$1.70 \pm 0.02$ & $2.15 \pm 0.03$ & 1.79 \\
40489 & 19492 & 1990 Dec 30 & $1.97 \pm 0.02$   & 1.68 &
$1.83 \pm 0.02$ & $2.34 \pm 0.03$ & 1.66 \\
42132 & 20891 & 1991 Jul 27 & $2.07 \pm 0.03$   & 1.40 &
$1.92 \pm 0.03$ & $1.54 \pm 0.02$ & 1.38 \\
44806 & 23207 & 1992 May 29 & $1.80 \pm 0.04$   & 1.82 &
$1.65 \pm 0.04$ & $1.51 \pm 0.02$ & 1.81 \\
56635 & 31908 & 1996 Jan 25 & $2.31 \pm 0.09$   & 1.49 & 
$2.16 \pm 0.09$ & $0.40 \pm 0.01$ & 1.48 \\
\hline
\multicolumn{8}{l}{$^a$ Results for both observed (col. 4) and
de-reddened spectra (cols. 6 and 7) are given}\\
\multicolumn{8}{l}{ ~~~ along with their respective reduced 
$\chi^2$ (cols. 5 and 8).}\\
\multicolumn{8}{l}{$^b$ De-reddened fitted flux at 2000 \AA. The  
interstellar extinction $A_V = 0.13$ implies,}\\
\multicolumn{8}{l}{ ~~~ at this wavelength, a correction of 
$\sim$40\% with respect to the observed flux.}\\
\multicolumn{8}{l}{Note. Errors represent 1-$\sigma$ uncertainties.}\\
\end{tabular}
\end{center}

 

\newpage

\begin{figure}
\epsfysize=15cm 
\hspace{3cm}\epsfbox{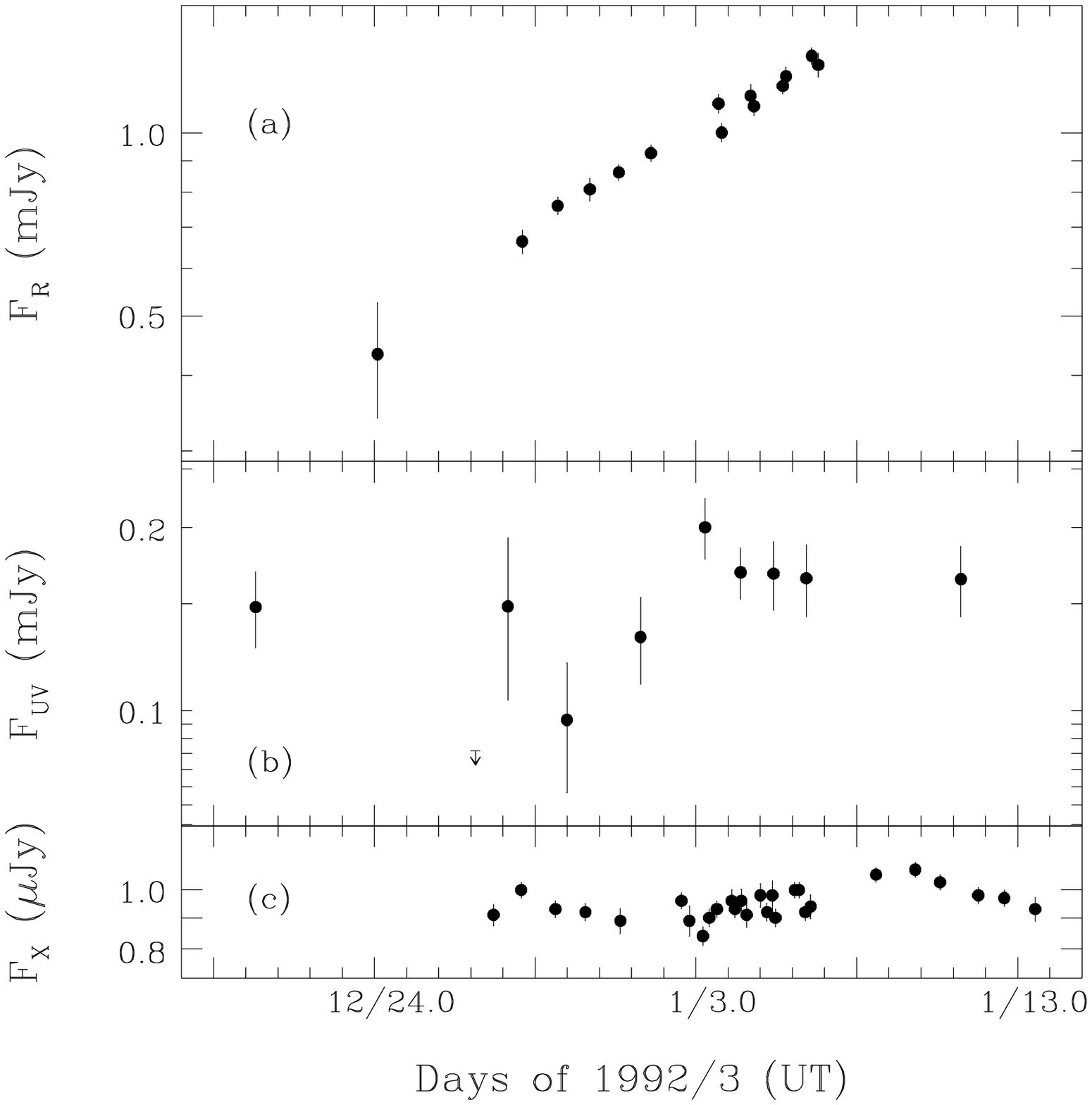} 
Fig. 1
\end{figure}

\newpage

\begin{figure}
\epsfysize=15cm 
\hspace{3cm}\epsfbox{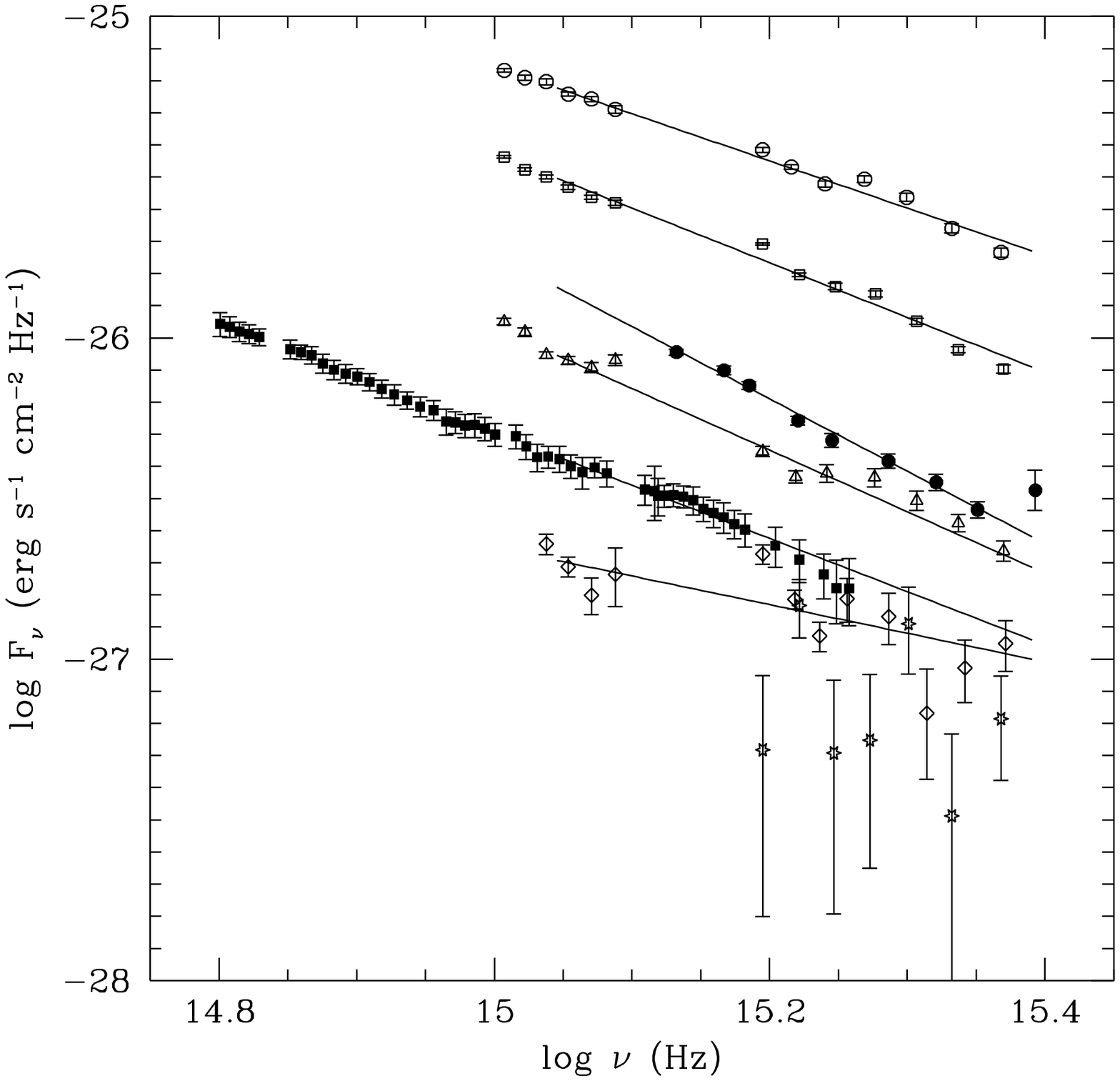} 
Fig. 2
\end{figure}

\newpage

\begin{figure}
\epsfysize=15cm 
\hspace{3cm}\epsfbox{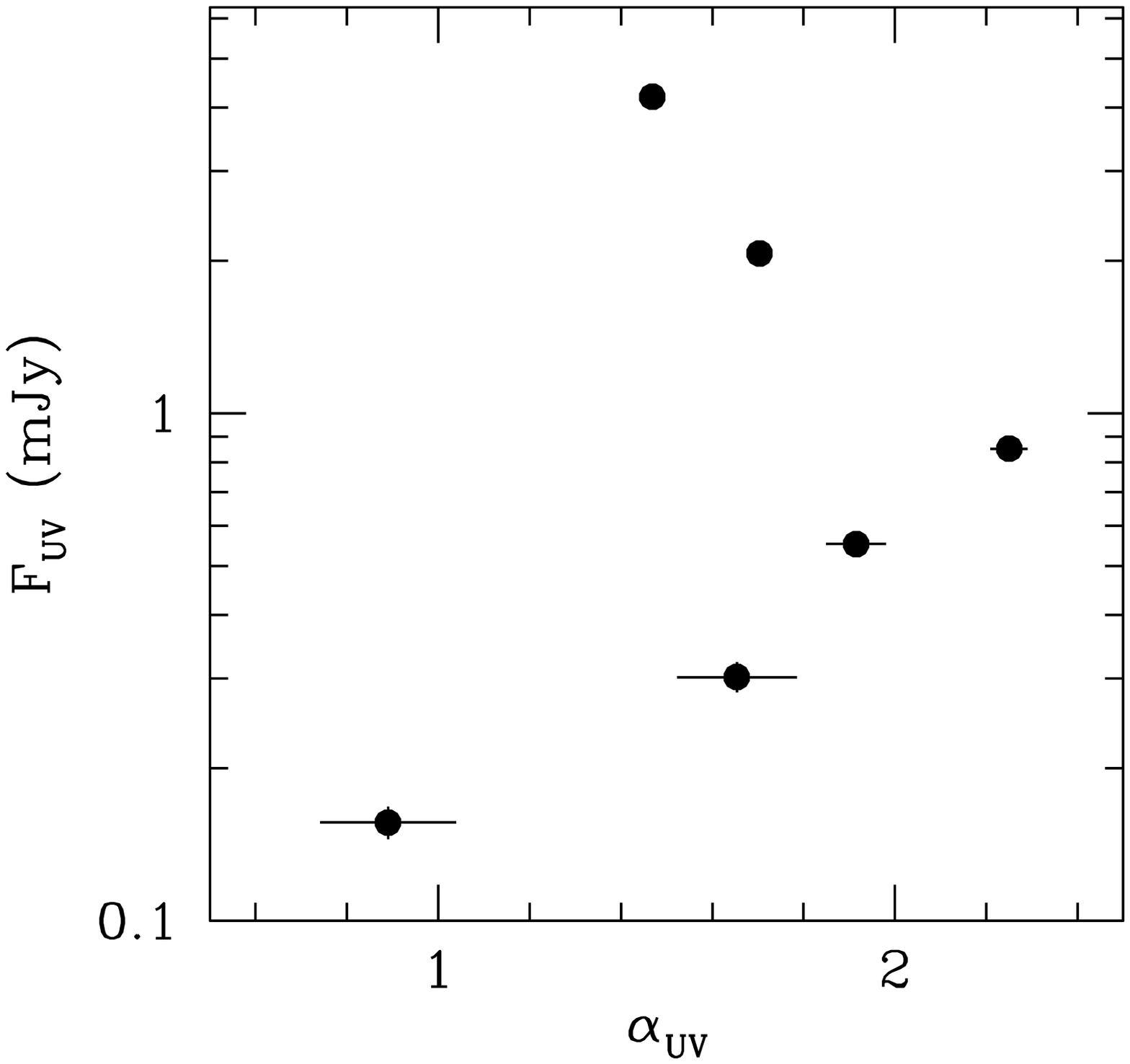} 
Fig. 3
\end{figure}

\newpage

\begin{figure}
\epsfysize=15cm 
\hspace{3cm}\epsfbox{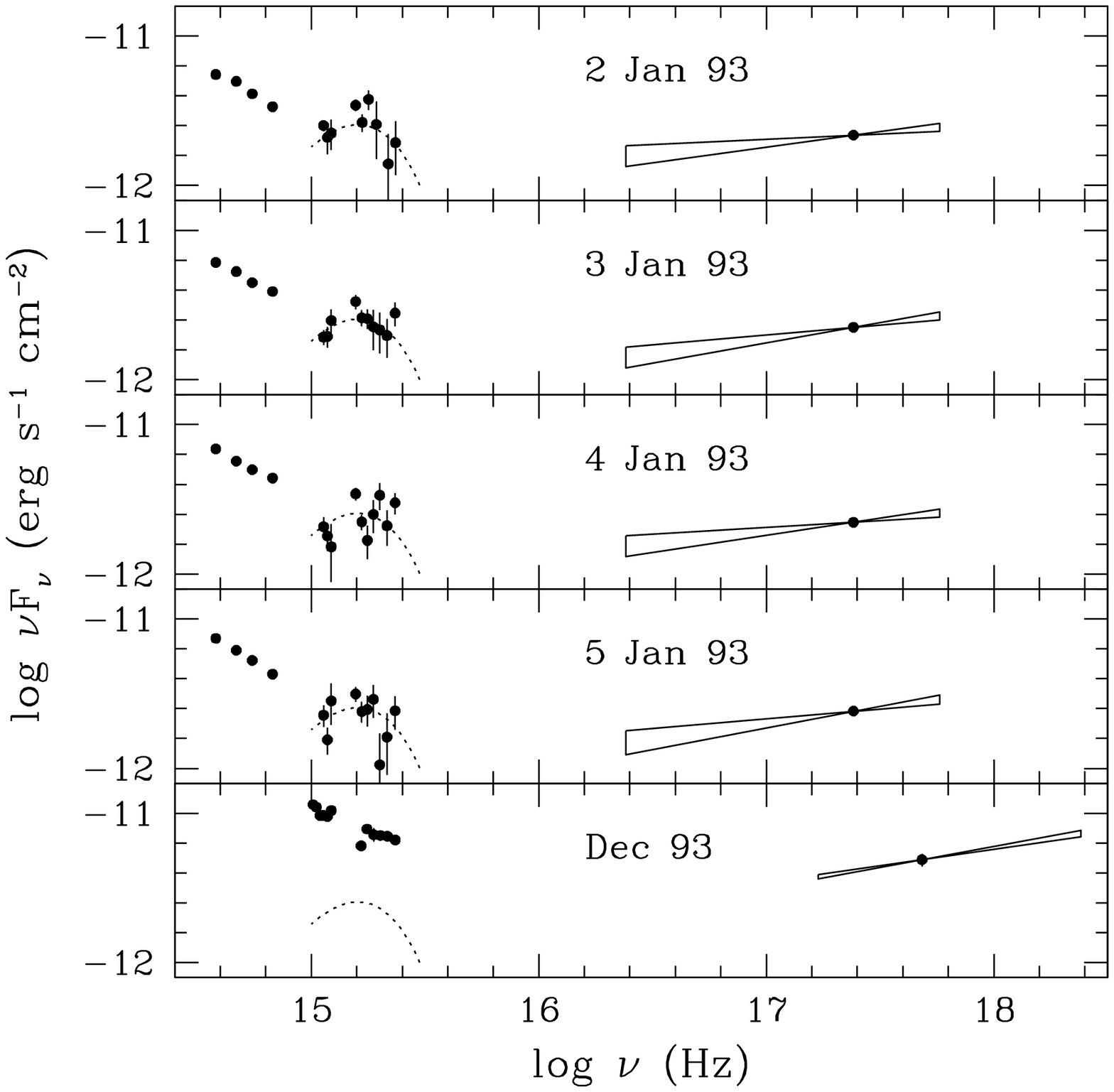} 
Fig. 4
\end{figure}

\newpage

\begin{figure}
\epsfysize=15cm 
\hspace{3cm}\epsfbox{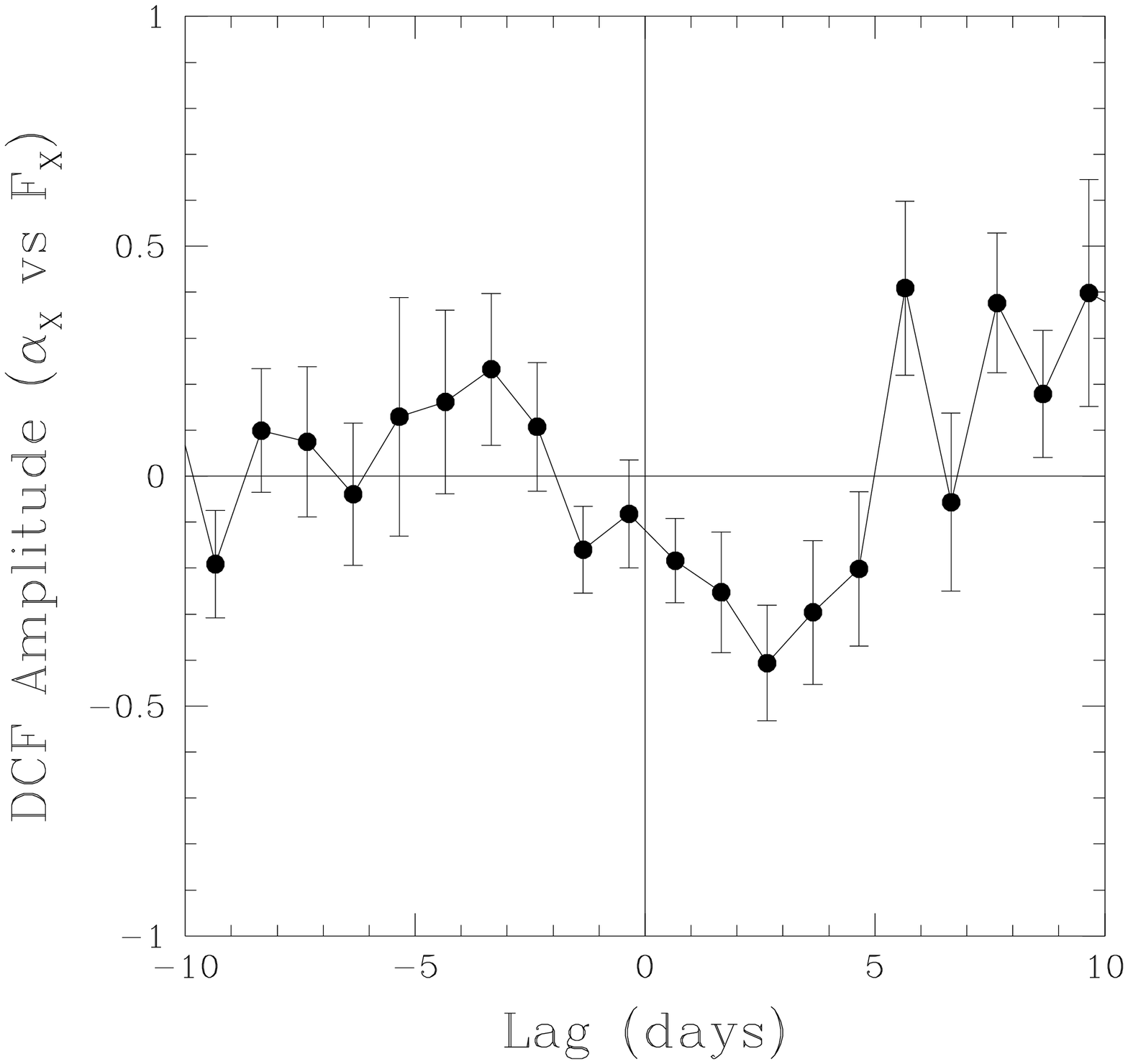} 
Fig. 5
\end{figure}

\newpage

\begin{figure}
\epsfysize=15cm 
\hspace{3cm}\epsfbox{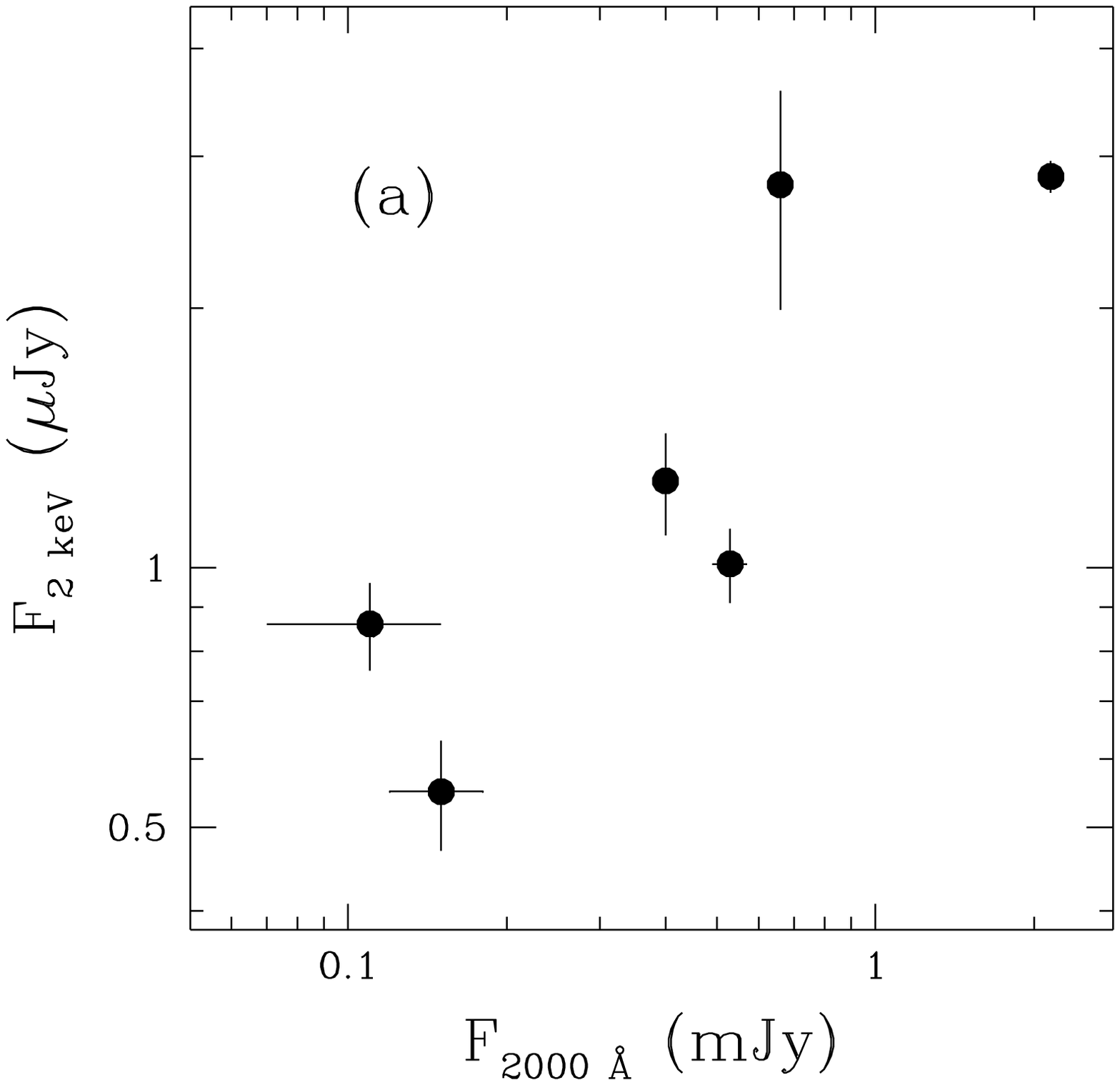} 
Fig. 6a
\end{figure}

\newpage

\begin{figure}
\epsfysize=15cm 
\hspace{3cm}\epsfbox{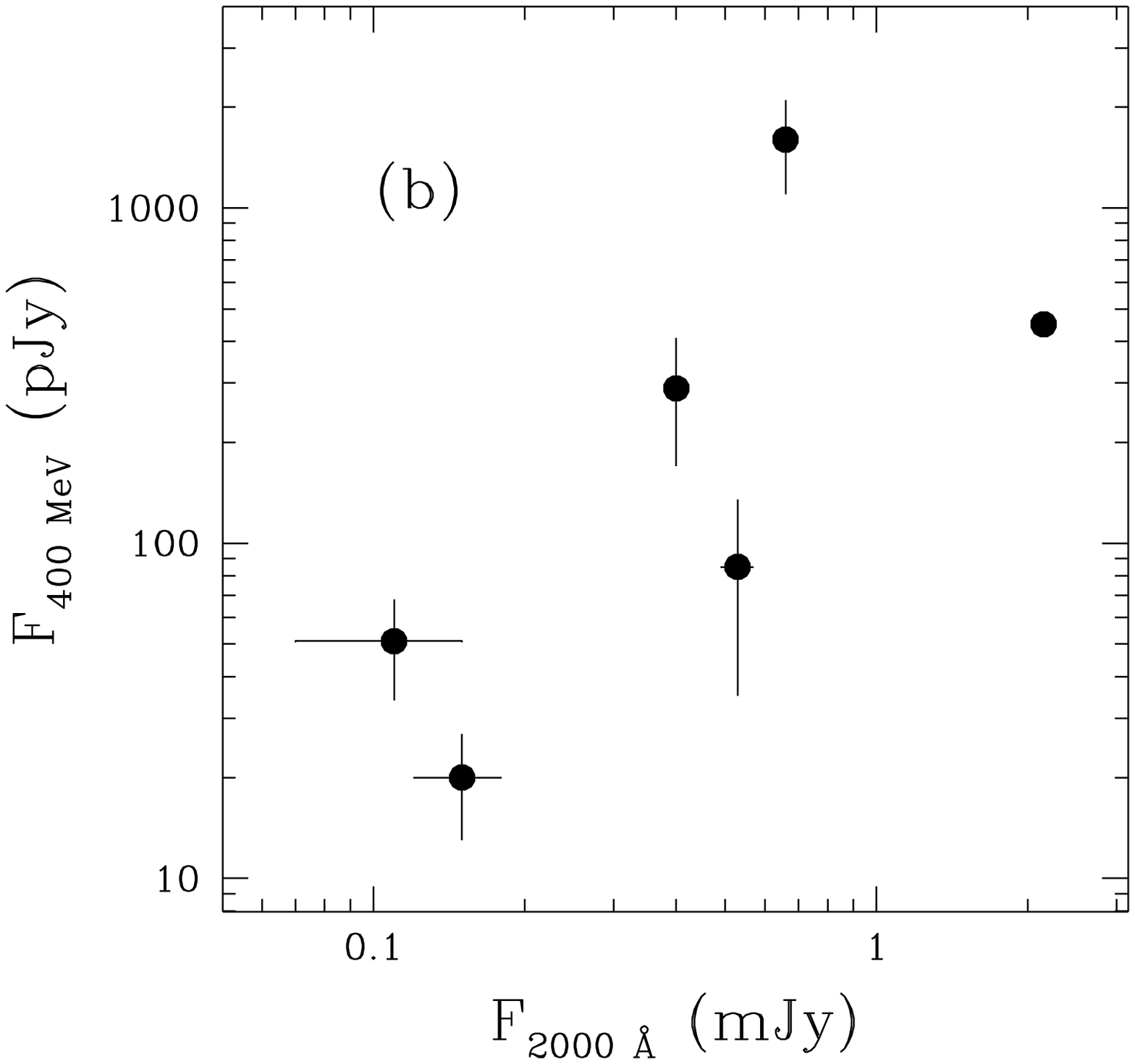} 
Fig. 6b
\end{figure}


\begin{references}

\ref{Aharonian, F., et al. 1997, A\&A, 327, L5}

\ref{Bloom, S. D., et al. 1997, ApJ, 490, L145}

\ref{Bonnell, J. T., Vestrand, W. T., \& Stacy, J. G. 1994, ApJ, 420, 545} 

\ref{Bregman, J. N. 1986, ApJ, 301, 708}

\ref{Bregman, J. N. 1990, A\&AR, 2, 125}

\ref{Bregman, J. N. 1994, in IAU Symp. 159, Multi-Wavelength Continuum
Emission of AGN, ed. T. J.-L. Courvoisier and A. Blecha (Dordrecht: Kluwer),
p. 5}


\ref{Brown, L. M. J., et al. 1989, ApJ, 340, 129}

\ref{Caplinger, J. 1995, NASA IUE Newsletter No. 55, 17}

\ref{Cardelli, J. A., Clayton, G. C., \& Mathis, J. S. 1989, ApJ, 345, 245}

\ref{Catanese, M., et al. 1997, ApJ, 487, L143}

\ref{Celotti, A., Maraschi, L., \& Treves, A. 1991, ApJ, 377, 403}

\ref{Chiappetti, L., et al. 1999, ApJ, 521, in press}

\ref{Corbett, E. A., Robinson, A., Axon, D. J., Hough, J. H., Jeffries,
 R. D., Thurston, M. R., \& Young, S. 1996, MNRAS, 281, 737}

\ref{Crenshaw, M. D., Bruegman, O. W., \& Norman, D. J. 1990, PASP, 102, 463}

\ref{Czerny, B., \& Elvis, M. 1987, ApJ, 321, 305}

\ref{Dermer, C., \& Schlickeiser, R. 1993, ApJ, 416, 458}

\ref{Edelson, R. A., \& Krolik, J. H. 1988, ApJ, 333, 646}

\ref{Edelson, R. A., Pike, G. F., Saken, J. M., Kinney, A., \& Shull, J. M.
1992, ApJS, 83, 1}


\ref{Elvis, M., Lockman, F. J., \& Wilkes, B. J. 1989, AJ, 97, 777}

\ref{Elvis, M., et al. 1994, ApJS, 95, 1}

\ref{Falomo, R., Treves, A., Chiappetti, L., Maraschi, L., Pian, E., \&
Tanzi, E. G. 1993, ApJ, 402, 532}

\ref{Gaidos, J. A., et al. 1996, Nature, 383, 319}

\ref{Garhart, M. P., Smith, M. A., Levay, K. L., \& Thompson, R. W. 1997, 
IUE NASA Newsletter No. 57, p.165}

\ref{Garhart, M. P. 1998, IUE NASA Newsletter No. 58, in press} 

\ref{Ghisellini, G., \& Maraschi, L. 1996, in ``Blazar Continuum
Variability", ASP Conf. Ser., Vol. 110, eds. H. R. Miller, J. R.
Webb, and J. C. Noble, p. 436}

\ref{Ghisellini, G., \& Madau, P. 1996, MNRAS, 280, 67}

\ref{Grandi, P., et al. 1996, ApJ, 459, 73}

\ref{Grandi, P., et al. 1997, A\&A, 325, L17}

\ref{Haardt, F., et al. 1998, A\&A, 340, 35}

\ref{Hartman, R. C., et al. 1992, ApJ, 385, L1}

\ref{Hartman, R. C., et al. 1996, ApJ, 461, 698}

\ref{Hartman, R. C. 1996, in ``Blazar Continuum Variability", ASP Conf. Ser.,
Vol. 110, eds. H. R. Miller, J. R. Webb, and J. C. Noble, p. 333}

\ref{Imhoff, C. 1996, IUE NASA Newsletter 56, 108}

\ref{Imhoff, C. 1997, IUE NASA Electronic Newsletter, Vol. 5, No. 4}

\ref{Kolman, M., Halpern, J. P., Shrader, C. R., Filippenko, A. V., 
Fink, H. H., \& Schaeidt, S. G. 1993, ApJ, 402, 514}

\ref{Koratkar, A., Pian, E., Urry, C. M., \& Pesce, J. E. 1998, ApJ,
492, 173}

\ref{Kubo, H., Takahashi, T., Madejski, G., Tashiro, M., Makino,
F., Inoue, S., \& Takahara, F. 1998, ApJ, 504, 693}

\ref{Lawson, A., \& McHardy, I. M. 1998, MNRAS, 300, 1023}

\ref{Loiseau, N., \& Schartel, N. 1998,
http://ines.vilspa.esa.es/ines/docs/contents.html}

\ref{Macomb, D. J., et al. 1995, ApJ, 449, L99}

\ref{Makino, F., et al. 1989, ApJ, 347, L9}

\ref{Maraschi, L., Chiappetti, L., Falomo, R., Garilli, B., Malkan, M.,
Tagliaferri, G., Tanzi, E. G., \& Treves, A. 1991, ApJ, 368, 138}

\ref{Maraschi, L., Ghisellini, G., \& Celotti, A. 1992, ApJ, 397, L5}

\ref{Maraschi, L., et al. 1994, ApJ, 435, L91}

\ref{Maraschi, L. 1998, Nucl. Phys. B (Proc. Suppl.), 69/1-3, 389}

\ref{Masnou, J. L., Wilkes, B. J., Elvis, M., McDowell, J. C., \&
Arnaud, K. A. 1992, A\&A, 253, 35}

\ref{Mattox, J. R., Wagner, S. J., Malkan, M., McGlynn, T. A.,
Schachter, J.  F., Grove, J. E., Johnson, W. N., \& Kurfess, J. D.
1997, ApJ, 476, 692}

\ref{Netzer, H. 1990, in Active Galactic Nuclei, Saas-Fee Advanced Course 
20, Lecture Notes 1990, Swiss Society for Astrophysics and Astronomy, eds.
T. J.-L. Courvoisier and M. Mayor (Springer Verlag)}

\ref{Netzer, H., et al. 1994, ApJ, 430, 191}

\ref{Nichols, J. S., \& Linsky, J. L. 1996, AJ, 111, 517}

\ref{Pian, E., et al. 1993, ApJ, 486, 784}

\ref{Rees, M. J. 1984, ARAA, 22, 471}

\ref{Rieke, G. H., \& Lebofsky, M. J. 1985, ApJ, 288, 618}

\ref{Sambruna, R. M. 1997, ApJ, 487, 536}


\ref{Schartel, N., Walter, R., Fink, H. H., \& Tr\"umper, J. 1996, A\&A,
307, 33}

\ref{Sembay, S., Warwick, R. S., Urry, C. M., Sokoloski, J., George, I. M.,
Makino, F., Ohashi, T., \& Tashiro, M. 1993, ApJ, 404, 112}

\ref{Shields, G. A. 1978, Nature, 272, 706}

\ref{Shrader, C. R., et al. 1994, AJ, 107, 904}

\ref{Shull, J. M., \& Van Steenberg, M. E. 1985, ApJ, 294, 599}

\ref{Sikora, M., Begelman, M., \& Rees, M. J. 1994, ApJ, 421, 153}

\ref{Takahashi, T., et al. 1996, ApJ, 470, L89}

\ref{Thompson, D. J., et al. 1996, ApJS, 107, 227}

\ref{Treves, A., et al. 1999, in ASP Conf. Ser. 159, BL Lac
Phenomenon, ed L. Takalo (San Francisco: ASP), p. 184}

\ref{Ulrich, M.-H., 1981, Space Sci. Rev., 28, 89}

\ref{Ulrich, M.-H., Maraschi, L., \& Urry, C. M. 1997, ARAA, 35, 445}

\ref{Urry, C. M., et al. 1993, ApJ, 411, 614}


\ref{Webb, J. R., Carini, M. T., Clements, S., Fajardo, S., Gombola, P. P.,
Leacock, R. J., Sadun, A., \& Smith, A. G. 1990, AJ, 100, 1452}

\ref{Wehrle, A. E., et al. 1998, ApJ, 497, 178}

\end{references}
\end{document}